\documentclass[twocolumn]{aastex631}

\usepackage{amsmath}
\usepackage{hyperref}

\usepackage{threeparttable}
\usepackage{makecell}

\begin{document}

\title{Observation of Wobbling and Precessing Jet Signatures in Gamma-Ray Emission from Blazar PKS 1424–41}

\author[0000-0002-5221-0822]{Ajay Sharma}
\altaffiliation{ajjjkhoj@gmail.com}
\affiliation{S. N. Bose National Centre for Basic Sciences, Block JD, Salt Lake, Kolkata 700106, India}

\author[0000-0003-1071-5854]{Debanjan Bose}
\altaffiliation{debanjan.tifr@gmail.com}
\affiliation{Department of Physics, Central University of Kashmir, Ganderbal, 191131, India}







\begin{abstract}


We report a clear observational evidence for jet precession and nutation manifested in the gamma-ray emissions of the blazar PKS~1424$-$41. An analysis of $\sim$16~yr of \emph{Fermi}-LAT observations reveals three distinct and superimposed quasi-periodic oscillations (QPOs), with characteristic timescales of $\sim$5.26~yr, $\sim$0.96~yr and $\sim$1.3~yr.The longer timescale is detected with a local significance of $\sim 4.1\sigma$ ($99.9958\%$ confidence) and a global significance of $\sim 2.45\sigma$ ($98.5714\%$ confidence). The shorter timescales ($\sim 0.96$ yr and $\sim 1.3$ yr) both show local significances exceeding $3\sigma$. To interpret the origin of these modulations, we examine two physically motivated scenarios: Lense-Thirring precession of a misaligned accretion disk and orbital-driven jet precession in a binary supermassive black hole (BSBH) system. We model the gamma-ray variability using a compound framework that incorporates jet precession together with intrinsic jet rotation (nutation), which successfully reproduces the observed year-like QPOs. Our results indicate that the long-term, persistent $\sim$5.26~yr oscillation is most consistent with a BSBH origin, while the shorter year-like QPO modulation can be naturally attributed to rotation or wobbling of the jet axis. If PKS~1424$-$41 hosts a supermassive black hole binary, it represents a promising target for future space-borne gravitational-wave observatories such as the Laser Interferometer Space Antenna (LISA), providing an independent probe of supermassive black hole binaries.

\end{abstract}


\section{Introduction} \label{sec:intro}
Blazar refers to a subclass of active galactic nuclei (AGNs) that includes BL Lacertae (BL Lac) and flat-spectrum radio quasars (FSRQ) objects. These sources are characterized by relativistic jets that are closely aligned with our line of sight, a configuration that has played a central role in shaping the current paradigm of AGN variability \citep{rees1978relativistic, begelman1980massive, ghisellini1993relativistic, urry1995unified, blandford2019relativistic}. The central engine is directly connected to the origin of temporal flux variability and to the emission observed across different wavebands throughout the electromagnetic spectrum \citep{urry1995unified, ulrich1997variability}. Numerous studies have reported the detection of QPOs across a wide range of electromagnetic wavelengths, including radio \citep{wang2014periodic, britzen2018oj287, tripathi2021quasi}, optical \citep{sandrinelli2016quasi}, X-rays \citep{wang2017nearly}, and $\gamma$-rays \citep{penil2020systematic, chen2024transient, banerjee2023detection, prince2023quasi, sharma2024detection, Sharma_Prince_Bose_2025, SHARMA2026100466}, spanning a broad range of characteristic timescales. Studies of jet variability provide a powerful means to probe the structure, physical conditions, and radiation mechanisms in AGN \citep{ulrich1997variability}. Among the phenomena revealed by such studies, quasi-periodic oscillations (QPOs) are particularly intriguing, albeit rare. The detection of year-like QPOs is especially uncommon and may point to the binary nature of the central supermassive black hole (SMBH) system, as suggested for well-known sources such as OJ~287 \citep{begelman1980massive, sillanpaa1988oj, lehto1996oj, katz1997precessing, valtonen2006predicting, valtonen2008massive} and PG~1553$+$113 \citep{caproni2017jet}, or alternatively to Lense-Thirring precession of the accretion disk \citep{romero2000beaming, britzen2018oj287, storchi1997evidence}.

The FSRQ PKS 1424$-$41 is one of the brightest known blazars, located at a redshift of $z = 1.522$ \citep{ajello20173fhl}. The blazar is highly variable FSRQ that has previously shown evidence of shorter-timescale quasi-periodic behaviour. Earlier studies reported transient or flare-associated periodicities of approximately ~94 days \citep{ren2023quasi}, ~355 days \citep{yang2021quasi}, and more recently transient periodicities of ~57 days and ~341 days \citep{chen2024transient} and 0.96 yr \citep{SHARMA2026100466}. However, all of these studies investigated either individual flaring episodes or transient periodic behaviour confined to limited time intervals, rather than searching for a persistent modulation over the complete Fermi-LAT monitoring period due to limited observation baseline.

Here, we report one of the longest quasi-periodic behaviour with timescales of $\sim$5.26~yr ever observed in the $\gamma$-ray emission of the blazar. In addition to this long-term modulation, we detect a shorter QPO with a timescale of $\sim$0.96~yr, providing evidence that the two periodic features likely originate from a binary supermassive black hole system and jet rotation about its own axis, respectively.
 Recently, \citep{robinson2024neutrino} estimated the black hole mass of PKS 1424$-$41 using a time-dependent lepto-hadronic model to fit its spectral energy distribution (SED), obtaining a mass of $M \sim 4 \times 10^{8}\,M_{\odot}$ and a bulk Lorentz factor of $\Gamma = 18$. The source exhibited pronounced flaring activity during the periods 2008-2014 \citep{celotti2008power, buson2014unusual, paliya2017general, abhir2021multi} and 2022-2023 \citep{SHARMA2026100452}. In addition to these major outbursts, PKS 1424$-$41 displayed recurrent sub-flaring events throughout the entire $\sim$16 yr observational span. Several studies have investigated the presence of quasi-periodic behavior in this source, particularly during the interval 2013-2019 \citep{bhatta2020nature, yang2021quasi, chen2024transient, ren2023quasi, SHARMA2026100466} and reported the QPO on period around 1 yr with significance exceeding $3\sigma$.

\section{Fermi-LAT DATA REDUCTION AND ANALYSIS} \label{sec:data}
 We analyzed the long-term $\gamma$-ray emission of PKS 1424–41 using data from the Fermi Large Area Telescope (LAT) covering the period from August 5, 2008 (MJD 54683) to April 5, 2024 (MJD 60405). The data reduction and likelihood analysis were carried out with the standard Fermi Science Tools (v11r05p3), following established procedures and applying all recommended event selection and background modeling criteria \citep{atwood2009large, cash1979parameter, mattox1996likelihood}. We constructed a 7-day binned light curve in the 100 MeV–300 GeV energy range, retaining only statistically significant detections with $\mathrm{TS} > 9$ \citep{prince2023quasi, sharma2024detection, Sharma_Prince_Bose_2025}. The gamma-ray light curve analysis was performed without incorporating the upper-limit measurements.

\begin{figure*}
    \centering
    \includegraphics[width=0.75\textwidth]{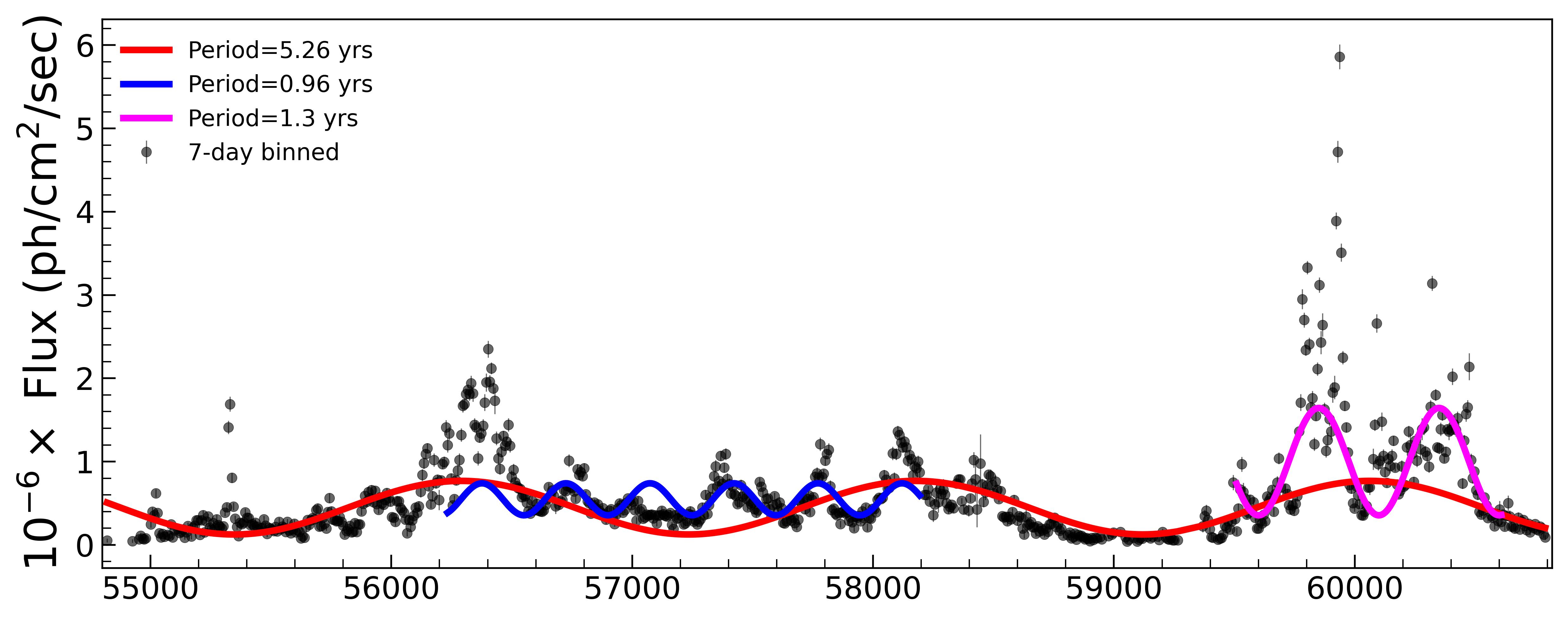}
    \caption{The 7-day binned Fermi-LAT gamma-ray light curve of PKS 1424-41, together with sinusoidal fits to the identified quasi-periodic modulations over different time intervals. The long-term QPO with a period of 5.26 yr is derived from the Lomb-Scargle periodogram (LSP) analysis of the complete 16-year light curve. A shorter QPO with a period of 0.96 yr is detected in the interval MJD 56254-58228, while another year-scale QPO with a period of 1.3 yr is identified during the later interval MJD 59509-60594. In the light curve, the upper limits were not plotted.}
    \label{fig-lc_sine}
\end{figure*}

\section{Search for Periodic Signatures} \label{sec:QPO_search}
To investigate possible periodic modulation in the $\gamma$-ray flux, we applied two independent and complementary time-series analysis techniques: the Lomb–Scargle periodogram (LSP) and the Weighted Wavelet Z-transform (WWZ) \citep{vanderplas2018understanding, banerjee2023detection, prince2023quasi, sharma2024detection,  Sharma_Prince_Bose_2025}. Both methods are well suited for unevenly sampled data and account for the influence of red-noise processes commonly present in AGN variability.

\begin{figure*}
    \centering
    \includegraphics[width=0.32\textwidth]{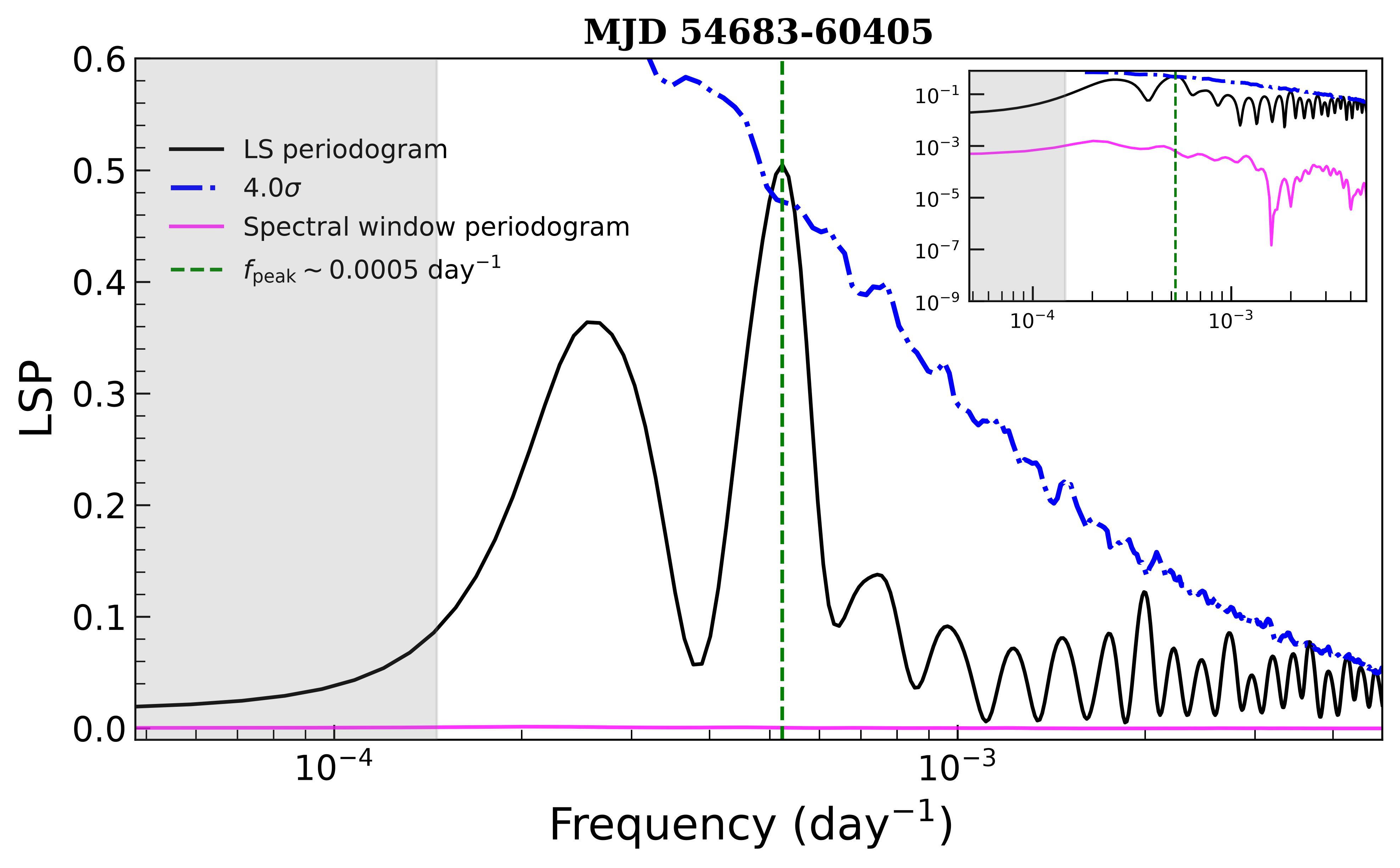}\hspace{1pt}
    \includegraphics[width=0.32\textwidth]{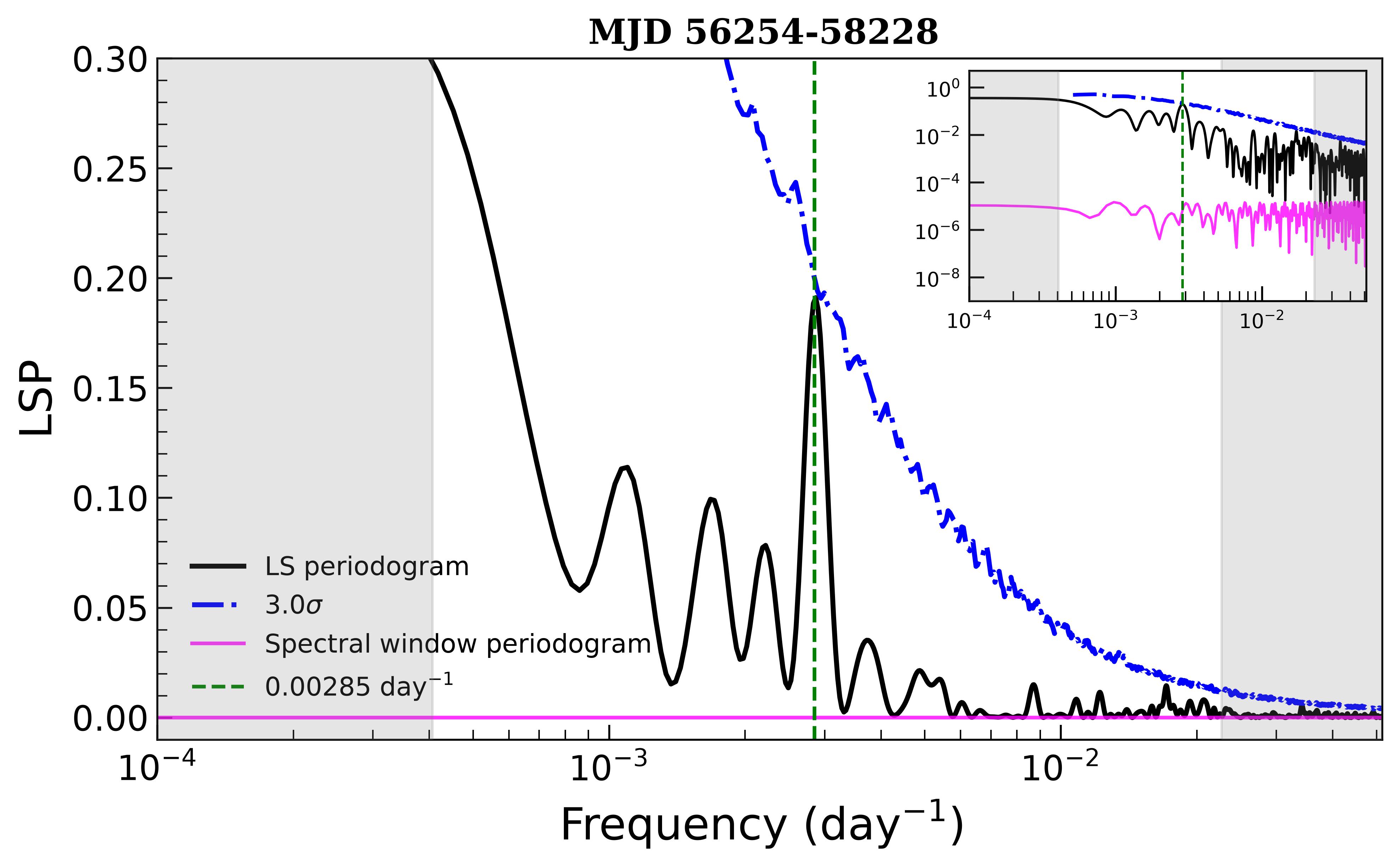}\hspace{1pt}
    \includegraphics[width=0.32\textwidth]{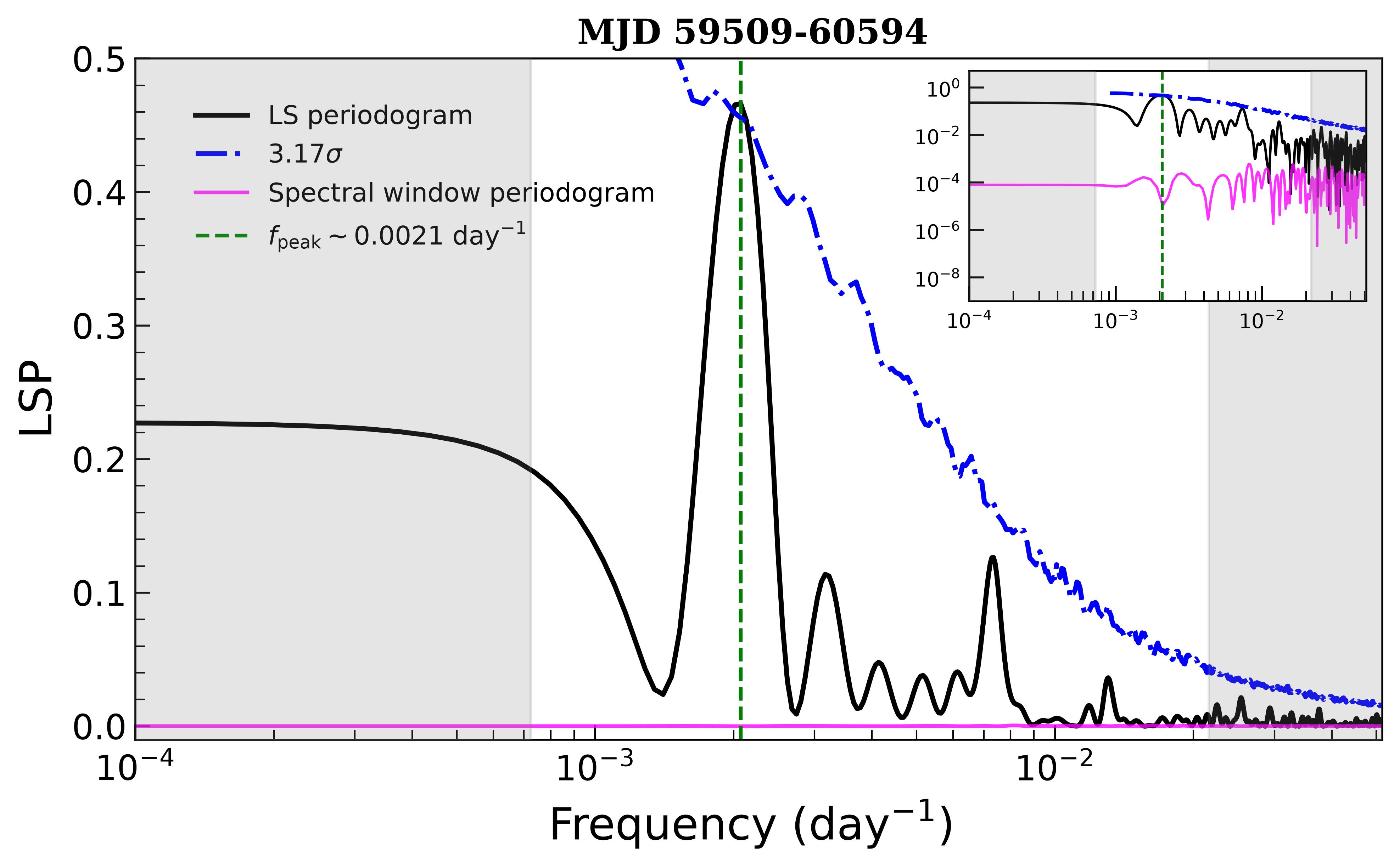}\vspace{1pt}
    \centering
    \includegraphics[width=0.32\textwidth]{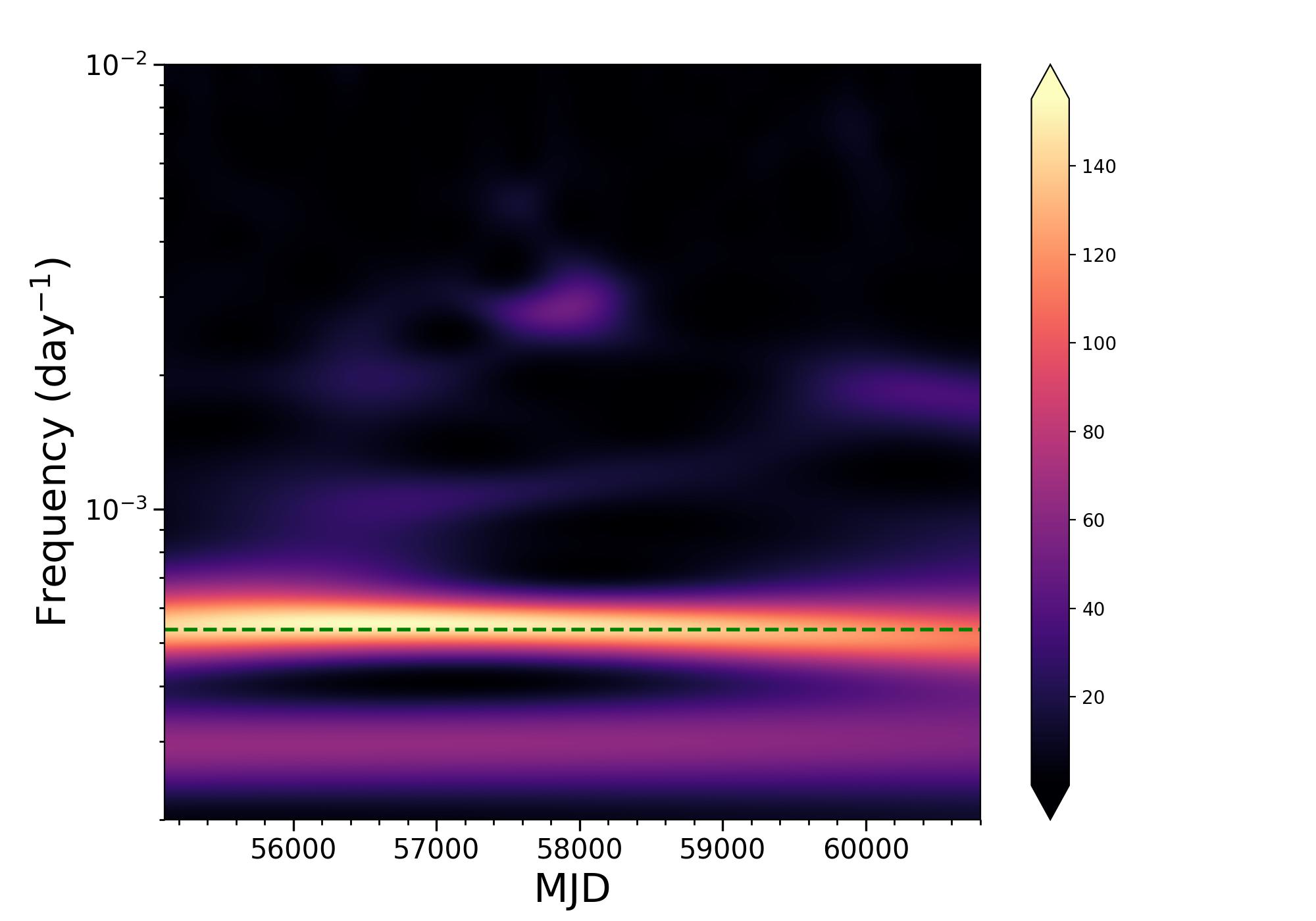}\hspace{1pt}
    \includegraphics[width=0.32\textwidth]{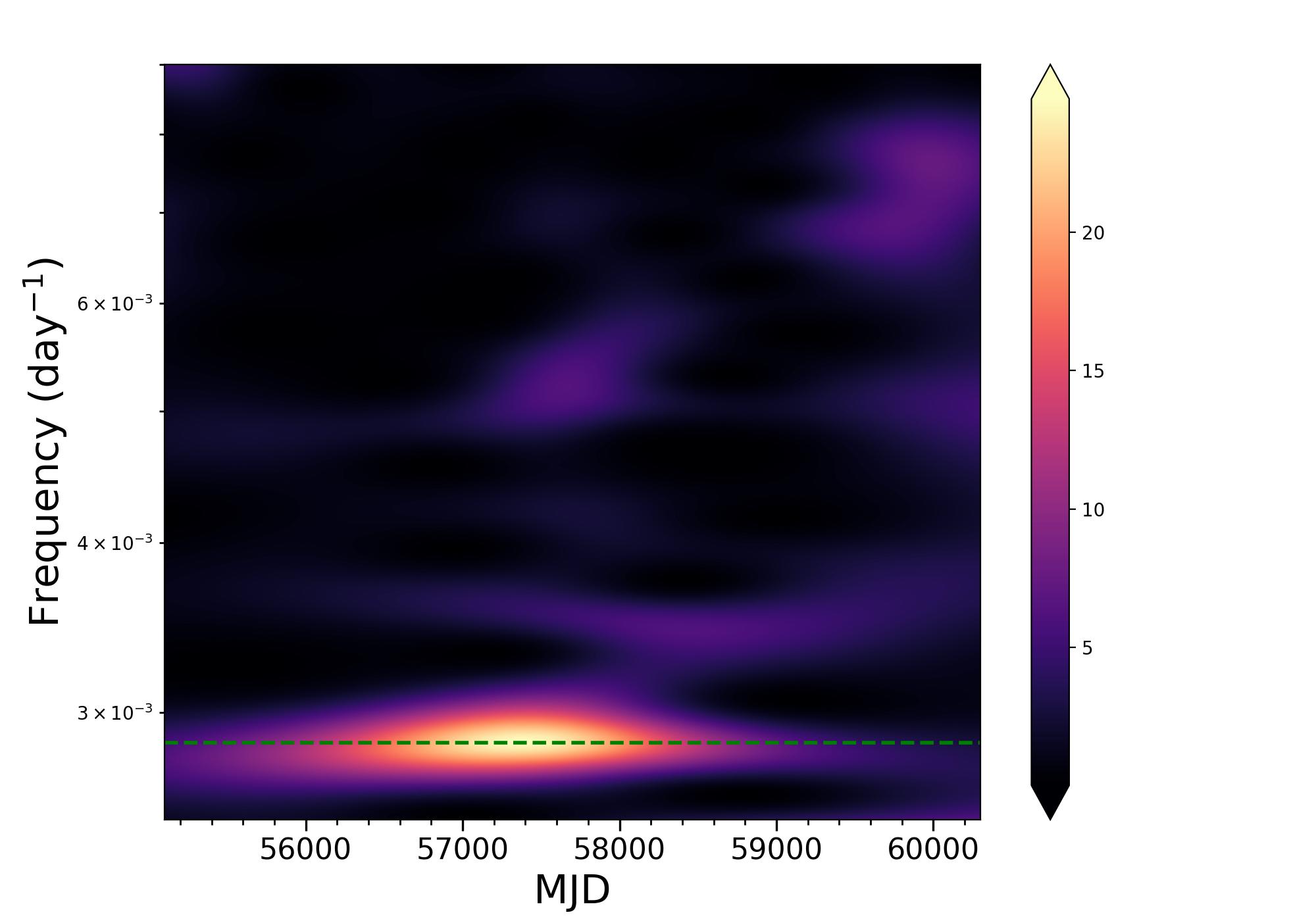}\hspace{1pt}
    \includegraphics[width=0.32\textwidth]{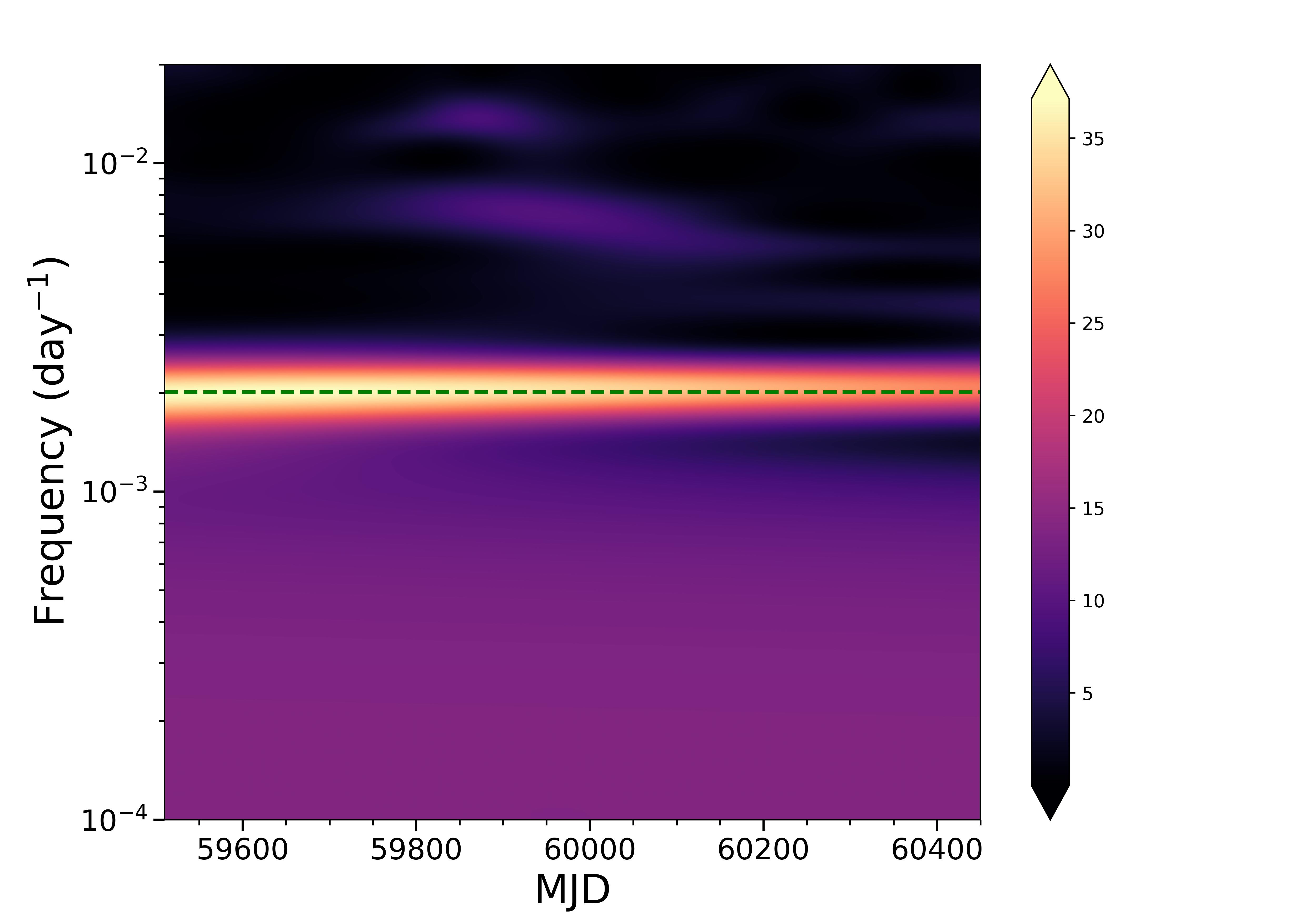}
    \caption{The LSPs of the 7-day binned $\gamma$-ray light curve of PKS~1424$-$41 are shown, with the black curve representing the LSP and the blue curve indicating the significance threshold. The top left panel shows the LSP corresponds to the full $\sim$16~yr dataset, while the top middle panel shows the LSP for the interval MJD~56254--58228 and the top right panel represents the LSP for the interval MJD~59509-60594. The bottom panels present the WWZ analyses with the color bar denotes the WWZ power, and the horizontal green dashed lines mark the detected QPO frequencies. Our study reveals these prominent QPO features at $\sim 5.3 \times 10^{-4}\,\mathrm{day^{-1}}$ (left), $\sim 2.85 \times 10^{-3}\,\mathrm{day^{-1}}$ (middle), and $\sim 2.1 \times 10^{-3}\,\mathrm{day^{-1}}$ (right) corresponding to the periods of 5.26 yr, 0.96 yr, and 1.3 yr, respectively. }
    \label{Fig-LSP_QPO}    
\end{figure*}

Our analysis reveals two statistically significant quasi-periodic signatures with characteristic timescales of $\sim 0.96 \pm 0.07$ yr, $\sim 1.3 \pm 0.23$ yr, and $\sim 5.26 \pm 0.72$ yr, detected at local significance levels exceeding $3\sigma$, $3\sigma$, and $4\sigma$, respectively, see Figures \ref{fig-lc_sine} and \ref{Fig-LSP_QPO}.\par


Active Galactic Nuclei (AGN) emission is intrinsically stochastic and is typically characterized by a red-noise process. When combined with irregular sampling, this variability can produce spurious peaks in periodograms that may mimic genuine periodic signals. Therefore, a rigorous statistical assessment is essential to evaluate the significance of any detected periodicity. To quantify the significance of the features identified in the Lomb-Scargle periodograms (LSPs), we followed the approach of \citep{emmanoulopoulos2013generating}. To fit the PSD of the light curve, we adopted the power-law model and performed the fitting using the DELightcurveSimulation code\footnote{\url{https://github.com/samconnolly/DELightcurveSimulation}}, a widely recognized method. This method model the underlying red-noise variability by fitting the power spectral density (PSD) and the probability density function (PDF) of the observed light curve. For the full $\gamma$-ray light curve, the best-fit PSD is described by a power-law model with a slope of $1.36$ and a normalization constant of $0.007535$. Using these parameters, we generated $2 \times 10^5$ simulated light curves that preserve the irregular sampling and statistical properties of the real data. For each realization, we computed an LSP using the identical procedure applied to the observed light curve. The distribution of LSP powers at each frequency was then constructed from the ensemble of simulations. The statistical significance of the observed peaks was determined by comparing them to the 84th, 97.5th, 99.85th, and 99.995th percentiles of these distributions, which correspond to the $1\sigma$, $2\sigma$, $3\sigma$, and $4\sigma$ confidence levels, respectively.


We computed the LSP for $2\times 10^5$ simulated light curves and counted the number of cases ($n_{\mathrm{pass}}$) in which the power at the target period of $\sim 5.26$ yr exceeds the observed peak power. We find $n_{\mathrm{pass}} = 0$, implying a local (single-frequency) p-value of $p \lesssim 10^{-5}$, corresponding to a local significance of $\sim 4.1\sigma$, which we refers to the periodicity detected at a single frequency in the light curve.\par

To account for the look-elsewhere effect, we adopted the approach from \citep{o2022unanticipated}, we identified the strongest peak in the LSP of each simulated light curve and computed its local p-value. We then counted the number of simulations ($n'_{\mathrm{pass}}$) in which this peak has a p-value equal to or lower than that of the observed $\sim 5.26$ yr signal. This yields $n'_{\mathrm{pass}} = 543$, corresponding to a global p-value of $5.58 \times 10^{-3}$, i.e., a global significance of $\sim 2.53\sigma$. While for shorter timescales, 3.0$\sigma$ (local significance) and 1.89$\sigma$ (global significance) for the 0.96 yr period, and sim 3.17$\sigma$ (local significance) and 2.13$\sigma$ (global significance) for the 1.3 yr period, see the Figure \ref{Fig-LSP_QPO}.

\section{Physical scenario} \label{sec:scenario}
\subsection{Jet precession}\label{sec:jet_precession}
We propose a simple precessing jet model to reproduce the approximate 5.26 year period modulation in the light curve of PKS 1424-41. With the reduced number of parameters, we obtain in this way a better constraint from the data. The precession model consists of a solid cone of half-opening angle $\Omega$, precessing with period $P_{\rm p}$ (angular velocity $\omega = 2\pi/P_{\rm p}$) about the precession cone axis directed along $z_{\rm p}$. We refer to Figure 11 in \citep{britzen2018oj287} for an equivalent geometrical scheme. The precession of an AGN jet is inherently a three-dimensional motion, yet astronomical observations provide only its two-dimensional projection on the plane of the sky. As a result, directly detecting variations in the jet inclination angle $\Phi$ is quite challenging. Instead, precession can be inferred indirectly by tracking changes in the jet position angle $\eta$, defined as the
angle between the projected jet direction and a chosen reference axis in the image plane. Both the inclination angle $\Phi$ and the position angle $\eta$ of the jet components can be expressed as functions of their projected (x,y) coordinates on the sky \citep{abraham2000precession, caproni2004can}, 

 \begin{align}
     \Phi (t) &= \mathrm{arcsin} \left( \sqrt{x(t)^2 + y(t)^2} \right), \nonumber \\
     \eta(t) &= \mathrm{arctan} \left( \frac{y(t)}{x(t)}\right),
     \label{eq:C3}
 \end{align}

\begin{align}
    x(t) &= A(t) \mathrm{cos}\eta_0 - B(t) \mathrm{sin}\eta_0 , \nonumber \\
    y(t) &= A(t) \mathrm{sin}\eta_0 + B(t) \mathrm{cos}\eta_0,
    \label{eq:C4}
\end{align}

 and with 

\begin{align}
    A(t) &= \mathrm{cos}\Omega \ \mathrm{sin} \Phi_0 + \mathrm{sin} \Omega \ \mathrm{cos} \Phi_0 \ \mathrm{sin} \omega(t-t_0), \nonumber \\
    B(t) &= \mathrm{sin} \Omega \ \mathrm{cos} \omega(t-t_0),
    \label{eq:C5}
\end{align}

where $\Phi_{0}$ is the angle between the cone axis and the line of sight and $\eta_{0}$ is the projected angle of the cone axis in the plane of the sky. The geometry of the model is schematically shown in Fig. \ref{fig-model}.

The changing Doppler factor is obtained by $\delta = \Gamma(1 - \beta \mathrm{cos} \Phi (t))^{-1}$, where $\Gamma = (1 - \beta^2)^{-1/2}$ is the bulk Lorentz factor and $\beta = \nu_{\rm jet}/c$ is the bulk velocity. We can obtain the varying observed flux due to the jet precession by substituting Eqs.~\ref{eq:C3}, \ref{eq:C4}, and \ref{eq:C5} into $F \propto \delta^3 F^{'}$ and the final expression is given as,

\begin{align}
    F(t) &\propto \frac{F^{'}}{\Gamma^3 \left[ 1 - \beta \mathrm{cos} \left( \mathrm{arcsin} \sqrt{x^2 + y^2} \right) \right]^3}
    \label{eq:C6}
\end{align}

The parameter $\eta_0$ is completely canceled out in Eq.~\ref{eq:C6}, indicating that $\eta_0$ does not affect the result.

\subsection{Jet precession and nutation}\label{sec:jet_precession_nutation}
By inspecting the outcome of Fourier-based analysis of the gamma-ray light curve, revealed an additional QPO feature with a period of approximately $\sim 0.96$ yr, in addition to the long-term modulation with period of $\sim5.26$ yr. This suggests the presence of an extra motion superimposed on the jet precession already identified. We hypothesize that this secondary periodicity may originate from the rotational motion of either a plasmoid like emitting component within the jet frame or wobbling of jet itself, as proposed by \citep{britzen2018oj287}. The intrinsic jet kinematics therefore offer valuable insight into the associated periodicity and the underlying physical nature of the system. In the frame of the jet, nutation motion can be expressed with the angular velocity $\omega_n = 2\pi/P_n$ as the vector components

\begin{align}
\mathrm{n}_{\mathrm{J}} &=
\begin{pmatrix}
x_{\mathrm{J}} \\
y_{\mathrm{J}} \\
z_{\mathrm{J}}
\end{pmatrix} 
=
\begin{pmatrix}
\sin\Omega_{\mathrm{n}} \ \sin[\omega_{\mathrm{n}}(t-t_0)] \\
\sin\Omega_{\mathrm{n}} \ \cos[\omega_{\mathrm{n}}(t-t_0)]  \\
\cos\Omega_{\mathrm{n}}
\end{pmatrix}
\end{align}

The transformation of this motion to the observer's frame using two rotation $R_z$ ad $R_y$. The nutation motion in the source frame can be calculated as

\begin{align}
    n_{\mathrm{s}} &= R_z (\omega_{\mathrm{p}}(t - t_0)) \ R_y (\Omega_{\mathrm{p}}) \ n_{\mathrm{J}} \nonumber \\
\end{align}

where,
\begin{align}
   R_z (\omega_{\mathrm{p}}(t - t_0))
    & = 
    \begin{pmatrix}
       \mathrm{cos} (\omega_\mathrm{p} (t - t_0)) & -\mathrm{sin} (\omega_\mathrm{p} (t - t_0)) & 0  \\
       \mathrm{sin} (\omega_\mathrm{p} (t - t_0)) & \mathrm{cos} (\omega_\mathrm{p} (t - t_0)) & 0 \\
       0 & 0 & 1
    \end{pmatrix}
\end{align}

\begin{align}
   R_y (\Omega_{\mathrm{p}})
    & = 
    \begin{pmatrix}
       \mathrm{cos} (\Omega_\mathrm{p} ) & 0 & \mathrm{sin} (\Omega_\mathrm{p} )  \\
       0 & 1 & 0 \\
       -\mathrm{sin} (\Omega_\mathrm{p}) & 0 & \mathrm{cos} (\Omega_\mathrm{p}) 
    \end{pmatrix}
\end{align}

To transform this nutation motion in the observer's frame, two more rotations are required. The vector component $n_0$ in the observer's frame can be calculated as

\begin{align}
    n_0 &= R_z (\eta_0) \ R_y (\Phi_0) \ n_{\mathrm{s}} \\ \nonumber
    & = 
    \begin{pmatrix}
       \mathrm{cos} (\eta_0) & -\mathrm{sin} (\eta_0) & 0  \\
       \mathrm{sin} (\eta_0) & \mathrm{cos} (\eta_0) & 0 \\
       0 & 0 & 1
    \end{pmatrix}
    \begin{pmatrix}
       \mathrm{cos} (\Phi_0 ) & 0 & \mathrm{sin} (\Phi_0)  \\
       0 & 1 & 0 \\
       -\mathrm{sin} (\Phi_0) & 0 & \mathrm{cos} (\Phi_0) 
    \end{pmatrix}
    \mathrm{n}_{\mathrm{s}}
\end{align}

\begin{figure*}
    \centering
    \includegraphics[width=0.7\textwidth]{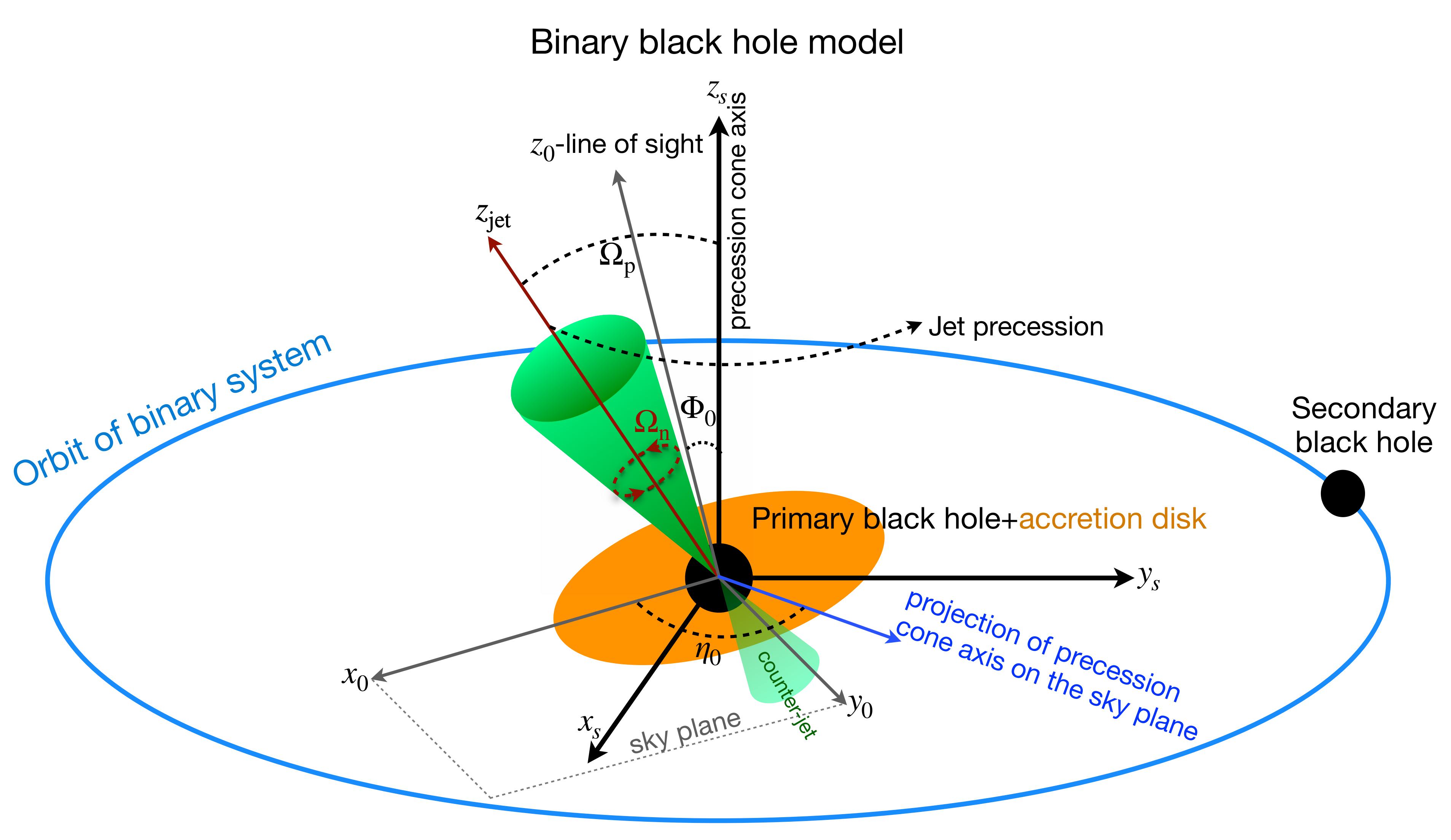}
    \caption{Schematic representation of the model describing the precession and rotation of the jet in PKS 1424–41. Three reference frames are illustrated: the jet frame, the source frame, and the observer frame. The half-opening angles of the precession cone ($\Omega_{\rm p}$) and the jet rotation $\Omega_{\rm n}$ are indicated. In the observer frame, the axis of the precession cone is characterized by the angles $\Phi_0$ and $\eta_0$.}
    \label{fig-model}
\end{figure*}

The combined geometry incorporating both the precession and nutation components of the jet is illustrated in Fig. \ref{fig-model}.

\section{Bayesian Inference for Model Fitting} \label{sec:LC_fitting}

We perform a likelihood analysis to model the $\gamma$-ray modulations with both models as mentioned above. To derive the final mathematical expression of our models, we used publicly available python library designed for symbolic mathematics, \textsf{sympy}\footnote{\url{https://www.sympy.org/en/index.html}} \citep{10.7717/peerj-cs.103}, and substituted in the expression of varying observed flux over time, which defined as $\propto \delta^3 F^{'}$. We consider a set of free parameters in the jet precession model ($\vartheta$= ($\Omega, P_{\rm p}, t_0, \Phi_0,\Gamma$)) and in the jet precession and nutation model ($\vartheta$= ($\Omega_{\rm p}, \Omega_{\rm n}, P_{\rm p}, P_{\rm n}, t_0, \Phi_0,\eta_0, \Gamma$)), and a Gaussian log-likelihood function 

\begin{align}
    \mathrm{ln}\mathcal{L}(\vartheta) = -\frac{1}{2} \sum_{i=1}^{\mathrm{n}} \left( \frac{ (Q_i^{\mathrm{obs}} - Q^{\mathrm{theor}}(t_{\rm i}, ~\vartheta))^2}{s_i^2} + \mathrm{ln} (2\pi s_i^2)\right),
    \label{eq:likelihood}
\end{align}

where $Q_i^{\mathrm{obs}}$ are observed flux, $Q^{\mathrm{theor}}(t_{\rm i}, ~ \vartheta)$ is the theoretical model flux, $s_i^2 = \sigma_{i,Q}^2 + \mathrm{exp}(2ln f)(Q^{\rm theor})^2$ is the variance, which contains an underestimation factor ln$f$ apart from the measurement error $\sigma_{i,Q}$. The model parameters were initially constrained by maximizing the posterior probability using the \textit{Nelder–Mead} simplex algorithm. Starting from physically motivated initial values, the optimization was performed by minimizing the negative log-posterior (equivalent to $\chi^2$). This step provided a stable best-fit solution that was subsequently used to initialize the MCMC sampling. We use the \textsf{emcee}\footnote{\url{https://emcee.readthedocs.io/en/stable/}} \citep{foreman2013emcee} MCMC module to sample from the posterior distribution. We set up 32 walkers and iterate over 50000 steps, discarding the first 5000 as burn-in to remove transient behavior. The posterior probability is given by Bayes’ theorem:

\begin{align}
    P(\vartheta|Q^{\rm obs}) = \frac{P(Q^{\rm obs}|\vartheta)P(\vartheta)}{P(Q^{\rm obs})}
    \label{eq:Bayes}
\end{align}

\begin{figure}
    \centering
    \includegraphics[width=0.49\textwidth]{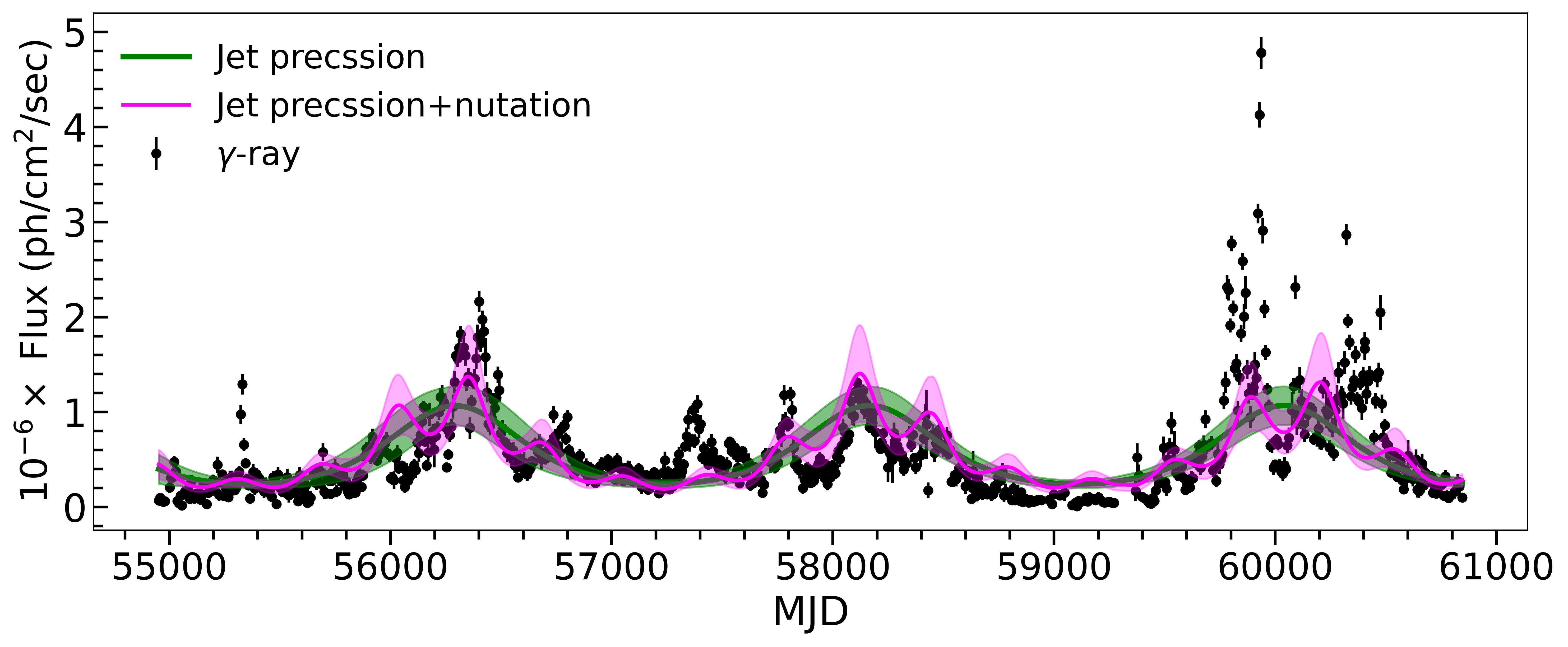}
    \caption{The 7-day binned $\gamma$-ray light curve of PKS 1424–41, shown by black data points, together with the best-fit jet-precession model (green curve) and a combined jet precession and nutation model (magenta curve). The shaded region represents the 3$\sigma$ uncertainty on the model prediction.}
    \label{fig-LC}
\end{figure}

\begin{figure*}
    \centering
    \includegraphics[width=0.6\textwidth]{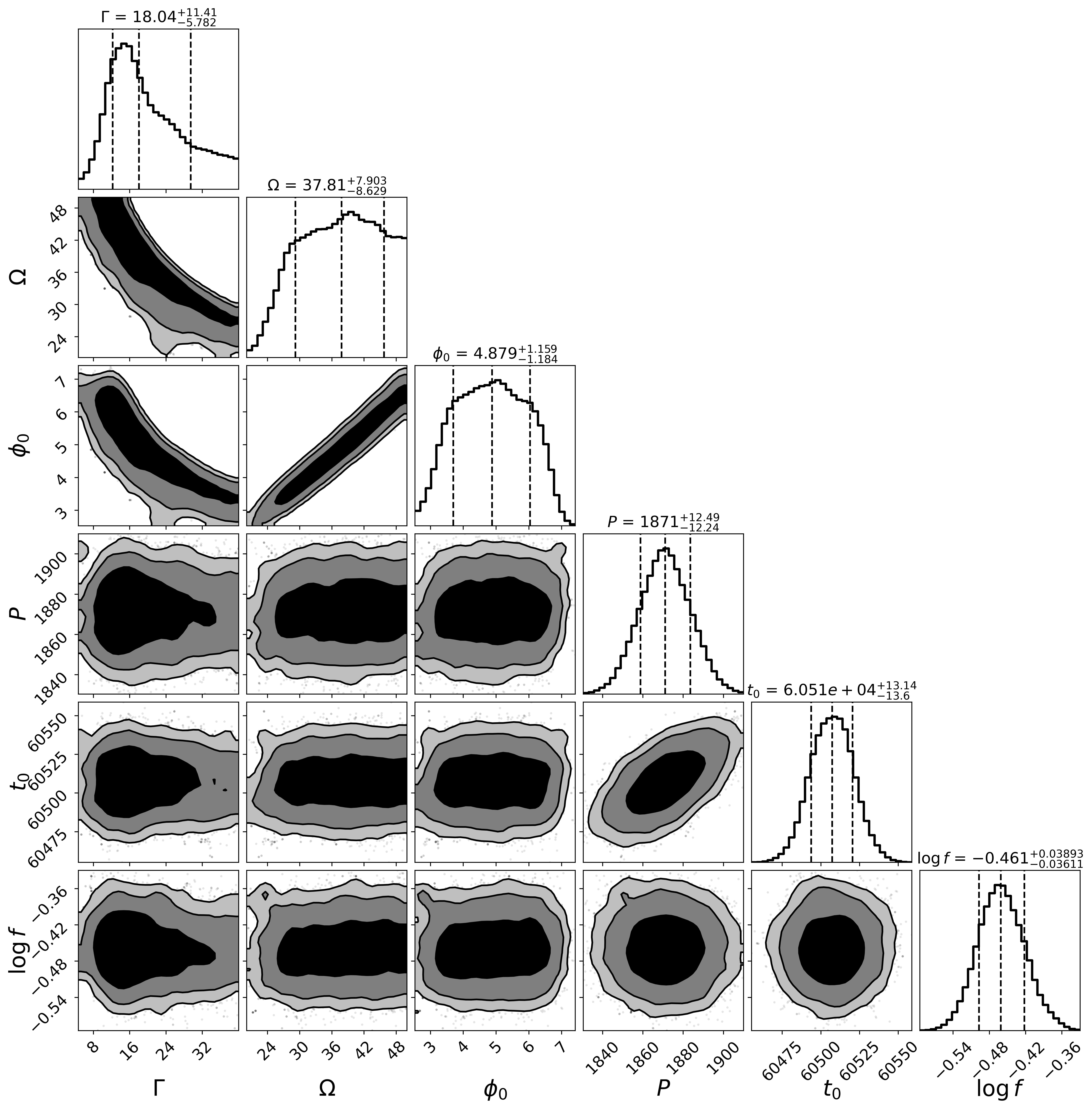}
    \caption{The posterior distribution of model parameters under the jet-precession framework. The 2D contours of all parameters show the 68, 95, and 99 per cent confidence levels estimated from MCMC analysis. The histograms show the projected 1D probability density distribution for all parameter with their best-fitting values and $1\sigma$ uncertainties. The normaliszation factor $F_0$ is estimated to be $2.8\times 10^{-5} \ \rm Ph \ \rm cm^{-2} \ \rm s^{-1}$ from the analysis .}
    \label{Fig-fitted_lc_jet_prec}    
\end{figure*}

\begin{figure*}
    \centering
    \includegraphics[width=0.9\textwidth]{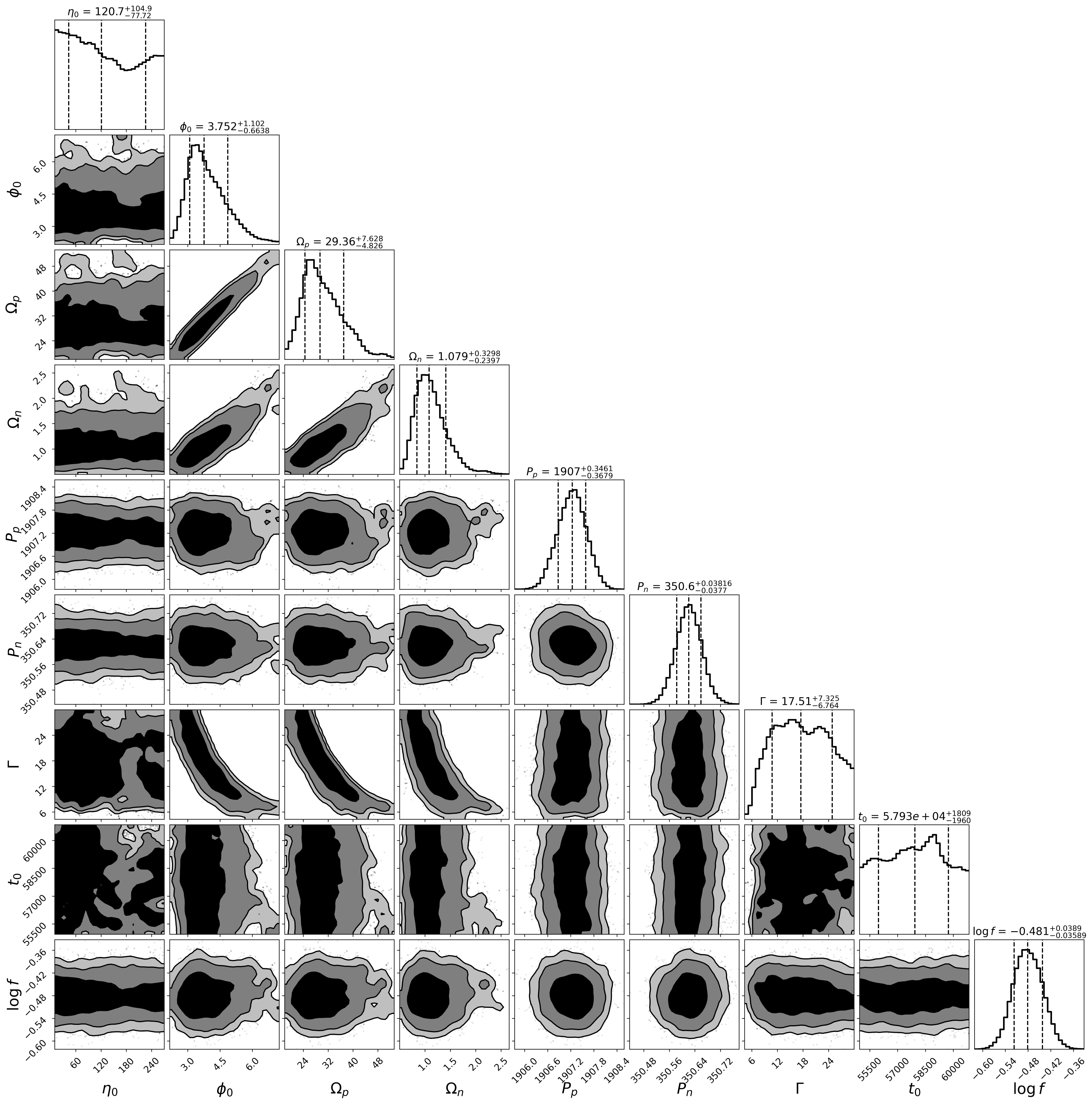}
    \caption{The posterior distribution of a jet-precession and nutation model parameters. The 2D contours of all parameters show the 68, 95, and 99 per cent confidence levels estimated from MCMC analysis. The histograms show the projected 1D probability density distribution for all parameter with their best-fitting values and $1\sigma$ uncertainties. The normaliszation factor $F_0$ is estimated to be $5.8\times 10^{-6} \ \rm Ph \ \rm cm^{-2} \ \rm s^{-1}$ from the analysis .}
    \label{Fig-corner_jet_prec_nut}    
\end{figure*}

The best-fit models parameters are summarized in Tables~\ref{tab:jet_precession} and \ref{tab:jet_precession_nutation} and Figures \ref{Fig-fitted_lc_jet_prec} and \ref{Fig-corner_jet_prec_nut}, and the fitted $\gamma$-ray light curve is shown in Fig. \ref{fig-LC}. Blazar emission arises from a combination of stochastic physical processes that are superimposed on some underlying physical mechanism, resulting in highly variable and often intricate light curves that are difficult to fully interpret. In this study, we model the $\gamma$-ray variability using a jet precession and nutation framework. While this model successfully reproduces the overall emission trend, it does not adequately capture the rapid and high-amplitude flux enhancements associated with flaring episodes, indicating that additional physical processes are likely contributing to the observed variability.

\begin{deluxetable}{ccc}
\setlength{\extrarowheight}{6pt}
\setlength{\tabcolsep}{10pt}
\tablecaption{Best-fit parameters of the jet precession model derived from the MCMC analysis, along with their uncertainties.
\label{tab:jet_precession}}
\tablehead{
\colhead{Parameter} & \colhead{Prior} & \colhead{Value} \\
(1) & (2) & (3)
}
\startdata
$t_0$ & 54683-60405 MJD & $60510_{-13.60}^{+13.14}$~MJD \\
$P_{\rm p}$ & 1000-2200 day & $1871_{-12.24}^{+12.49}$~day \\
$\Gamma$ & 1-40 & $18.04_{-5.78}^{+11.41}$ \\
$\Omega$ & 1-70$^{\circ}$ & $37.81_{-8.62}^{+7.90}\,^{\circ}$ \\ 
$\Phi_0$ & 0.1 - 10$^{\circ}$ & $4.87_{-1.18}^{+1.15}\,^{\circ}$ \\
$\log f$ & $[-2,\, +2]$ & $-0.46_{-0.036}^{+0.038}$ \\
\enddata
\begin{tablenotes}
\item Note: Column (1):  $t_0$ is reference time, $P_{\rm p}$ is the precession period, $\Gamma$ is the Lorentz factor, $\Omega$ and $\Phi_0$ are half-opening angle of the precession cone and angle between the precession cone and the line of sight, and correction factor (log $f$). Column (2) includes the prior range for each parameter and best optimal parameter values are summarized in column (3).
\end{tablenotes}
\end{deluxetable}


\begin{deluxetable}{ccc}
\setlength{\extrarowheight}{6pt}
\setlength{\tabcolsep}{10pt}
\tablecaption{Best-fit parameters of the jet precession-nutation model and their associated uncertainties derived from the MCMC analysis.
\label{tab:jet_precession_nutation}}
\tablehead{
\colhead{Parameter} & \colhead{Prior} & \colhead{Value} \\
(1) & (2) & (3)
}
\startdata
$t_0$ & 54683-60405 MJD & $57930_{-1960}^{+1809}$~MJD \\
$P_{\rm p}$ & 1000-2200 day & $1907_{-0.36}^{+0.34}$~day \\
$P_{\rm n}$ & 100-500 day & $350_{-0.037}^{+0.038}$~day \\
$\Gamma$ & 1-40 & $17.51_{-6.76}^{+7.32}$ \\
$\Omega$ & 1-50$^{\circ}$ & $29.36_{-4.82}^{+7.62}\,^{\circ}$ \\ 
$\Omega_{\rm n}$ & 1-8$^{\circ}$ & $1.079_{-0.23}^{+0.32}\,^{\circ}$ \\
$\Phi_0$ & 0.1 - 10$^{\circ}$ & $3.75_{-0.66}^{+1.10}\,^{\circ}$ \\
$\eta_0$ & 0 - 360$^{\circ}$ & $120_{-77.72}^{+104.9}\,^{\circ}$ \\
$\log f$ & $[-2,\, +2]$ & $-0.48_{-0.035}^{+0.038}$ \\
\enddata
\begin{tablenotes}
\item Note: Column (1):  $t_0$ is reference time, $P_{\rm p}$ is the precession period, $P_{\rm n}$ is the nutation period, $\Gamma$ is the Lorentz factor, $\Omega$ and $\Omega_{\rm n}$ are half-opening angle of the precession cone and jet, $\Phi_0$ is angle between the precession cone and the line of sight, $\eta_0$ is the projected angle of the cone axis in the plane of the sky, and correction factor (log $f$). Column (2) includes the prior range for each parameter and best optimal parameter values are summarized in column (3).
\end{tablenotes}
\end{deluxetable}

\section{Physical interpretation}\label{sec:physical_interpretation}
AGN jet precession may arise from various physical scenarios, such as a misaligned supermassive binary black hole (SMBBH) system or Lense–Thirring (LT) precession \citep{lense1918einfluss, thirring1918effect, abraham2018jet}. In the following, we examine these scenarios as separate case studies.


\subsection{Lense-Thirring precession of the BH}\label{sec:Lense_BH}
In a system hosting a single rotating (Kerr) black hole, the spin axis of the black hole can be misaligned with the angular momentum of the accretion disk, as suggested by the observed jet orientation \citep{caproni2004can}. This misalignment induces Lense–Thirring (LT) precession through frame-dragging, causing the inner accretion disk, and possibly the black hole itself, to precess. If the outer disk carries sufficient angular momentum, it can torque the inner regions, resulting in coherent precession of the disk–jet system. The precession period $P_{\mathrm{LT}}$ of the BH as a function of the rotation parameter $a_{*}$, the viscosity parameter $\alpha_{\mathrm{vis}}$, the BH mass $M$, and accretion rate $\dot{M}$ from \citep{ju1992alpha}, 

\begin{align}
    P_{\mathrm{LT}} &= 2\pi / \Omega_{\mathrm{LT}} ~~~ (\rm yr) \nonumber \\ 
    &= 10^{9.25} a_{*}^{0.71} \alpha_{\mathrm{vis}}^{1.37} \left( \frac{M}{10^8 \mathrm{M}_{\odot}}\right)^{1/7} \left( \frac{\dot{M}}{10^{-2} \mathrm{M}_{\odot} \mathrm{yr}^{-1}}\right)^{-6/5}
    \label{eq:LT_acc}
\end{align}

For a Kerr spin parameter $a_{*} = 0.1$ and a viscosity parameter $\alpha_{\mathrm{vis}} = 0.1$, adopting the formulation of \citep{king2007accretion} yields a Lense–Thirring precession period of $P_{\mathrm{LT}} \sim 1945\ \mathrm{yr}$ for a black hole mass of $M = 4\times 10^8 \ \mathrm{M}{\odot}$ adopted from \citep{robinson2024neutrino}, with an accretion rate of $\dot{M} \sim 20\ \mathrm{M}{\odot}\ \mathrm{yr}^{-1}$ \citep{jolley2009accretion}. Even for a higher accretion rate of $\dot{M} \sim 200\ \mathrm{M}{\odot}\ \mathrm{yr}^{-1}$ \citep{jolley2009accretion}, the corresponding precession period decreases only to $P_{\mathrm{LT}} \sim 122\ \mathrm{yr}$. To achieve the required Lense-Thirring precession period, $P_{\rm LT} = P_{\rm obs}/(1+z) \simeq 2.08~{\rm yr}$, one would require either an extremely small black hole spin parameter, of order $\sim 10^{-19}$, or a high mass accretion rate, $\dot{M} \sim 5.9\times10^{3}\,M_{\odot}\,{\rm yr}^{-1}$, which looks unrealistically. In this scenario, the LT precession timescale is expected to be significantly longer than the observed variability timescale of a few years, indicating that black hole precession driven by this mechanism is unlikely to explain the observed variability.

\subsection{Spin-induced precession in accretion disc}\label{sec:Lense_disc}
\begin{figure*}
    \centering
    \includegraphics[width=0.75\textwidth]{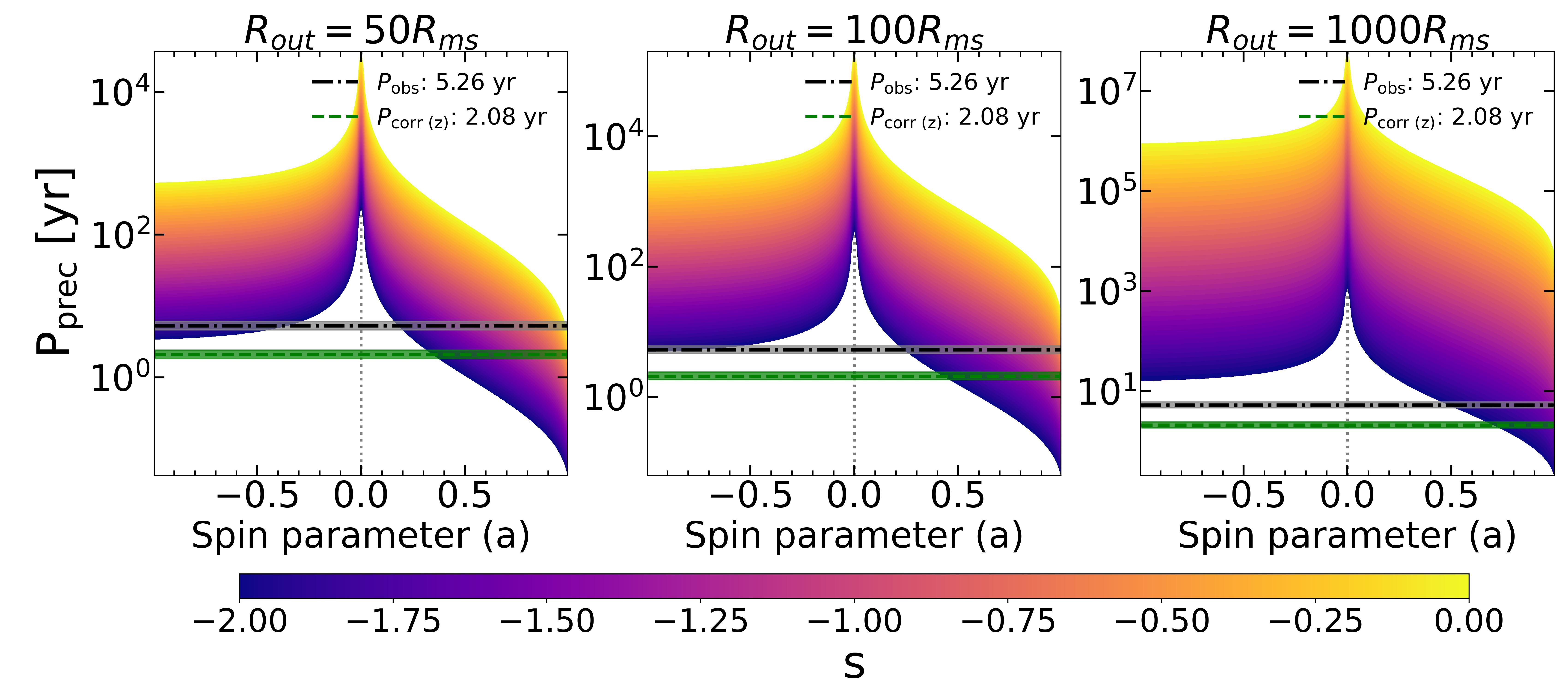}
    \caption{Precession period of PKS 1424–41 as a function of the black hole spin parameter. In the left panel, the outer radius of the accretion disk is fixed at $R_{\rm out} = 50 R_{\rm ms}$. The color bar indicates different power-law indices of the disk surface density, spanning the range $-2 < s < 0 $. The horizontal black and green lines represent the observed and redshift-corrected precession periods, respectively, along with their associated uncertainties. The middle and right panels show the same quantities as the left panel, but for disk outer radii of $R_{\rm out}=100 R_{\rm ms}$ and $R_{\rm out}=1000 R_{\rm ms}$, respectively. This analysis indicates that the inferred black hole spin is prograde.}
    \label{fig-LT_Precession_Disc}
\end{figure*}

If the accretion disk undergoes rigid-body precession, the Lense-Thirring (LT) effect is significant only in the inner regions, since the LT angular frequency decreases rapidly with radius \citep{wilkins1972bound}. The precession period of the disk–jet system primarily depends on the black hole angular momentum, $ J_{\mathrm{BH}} = G M_{\mathrm{BH}}^2 |a_{*}|/c$, where $a_{*}$ is the dimensionless Kerr spin parameter ($-1 \le a_{*} \le 1$), with positive (negative) values corresponding to prograde (retrograde) rotation. The precessing portion of the disk extends from $R_{\mathrm{in}} = \xi_{\mathrm{in}} R_{\mathrm{g}}$ to $R_{\mathrm{out}} = \xi_{\mathrm{out}} R_{\mathrm{g}}$, where $R_{\mathrm{g}} = GM_{\mathrm{BH}}/c^{2}$. The inner radius is commonly associated with the marginally stable orbit, $R_{\mathrm{ms}} = \xi_{\mathrm{ms}} R_{\mathrm{g}}$, which strongly depends on spin ($\xi_{\mathrm{ms}} = 1$ for maximally prograde and 9 for maximally retrograde rotation) \citep{caproni2004observational}. The LT precession period also depends on the disk surface density profile, typically assumed as a power law $\sum_{\mathrm{d}} = \sum_{\mathrm{0}} \xi^s$, with $s$ ranging from -2 to 0 \citep{papaloizou1995dynamics, larwood1997tidal, nelson2000hydrodynamic}. Under these assumptions, the precession period is obtained by integrating over the precessing disk region following \citep{caproni2004observational}.

\begin{align}
    P_{\mathrm{prec}} &= \frac{2\pi G M}{C^3} \ \frac{\int_{\xi_{\mathrm{ms}}}^{{\xi_{\mathrm{out}}}} \ \sum_{\mathrm{d}} (\xi) \ [\Upsilon (\xi)]^{-1} \ \xi^3 \mathrm{d}\xi}{\int_{\xi_{\mathrm{ms}}}^{{\xi_{\mathrm{out}}}} \ \sum_{\mathrm{d}} (\xi) \ \Psi (\xi) \ [\Upsilon (\xi)]^{-2} \ \xi^3 \mathrm{d}\xi} \ ,
    \label{eq:disk_prec}
\end{align}

where $\Upsilon (\xi) = \xi^{3/2} + a_{*}$ and $\Psi(\xi) = 1 - (1 - 4a_{*}\xi^{-3/2} + 3 a_{*}^2 \xi^{-2})^{/1/2}$. Applying this framework to PKS 1424-41, assuming outer disk radii of $R_{\mathrm{out}} = 50, 100, \ \mathrm{and}\ 1000 R_{\mathrm{ms}}$, the predicted LT precession period was computed as a function of the spin parameter $a_{*}$, as shown in Fig~\ref{fig-LT_Precession_Disc}.The observed jet precession period of PKS 1424–41, $P_{\mathrm{obs}} \simeq 5.26$ yr (corresponding to $\sim 2.08$ yrin the source frame), can be reproduced within the Lense–Thirring framework for a range of disk surface density profiles. For the adopted outer radii, the observed period is achievable only in the prograde configuration (see Fig.~\ref{fig-LT_Precession_Disc}). In the case of a steep density profile ($s = -2$), the period can be matched with moderate black hole spins of $a_* \sim 0.3, 0.35$ and 0.68 for outer radii of 50, 100, and 1000 $R_{\mathrm{ms}}$, respectively. Shallower profiles ($s = -1$)require comparatively higher prograde spins, particularly for larger outer radii. Although the results are sensitive to poorly constrained parameters such as the disk surface density profile, outer disk radius, and black hole spin, our analysis demonstrates that Lense–Thirring precession can account for the observed period. This supports a scenario in which the jet precession in PKS 1424–41 is driven by LT-induced disk precession around a single rotating black hole, consistent with precession timescales reported for other AGN jets \citep{britzen2018oj287}.

\subsection{A Supermassive Binary Black Hole Scenario}\label{sec:SMBBH}
Jet precession in PKS 1424--41 can be driven by gravitational torques exerted on a misaligned accretion disk by a secondary supermassive black hole in a binary system (SMBBH). The resulting disk-jet precession naturally produces year-scale oscillations, consistent with those observed in OJ 287 and PG 1553+113 \citep{britzen2018oj287, katz1997precessing, caproni2004can}. The binary orbital period obeys Kepler's third law: $r_{\rm ps}^3 = \frac{G M_{\rm tot}}{4\pi^2}\,P_{\rm ps}^2$, where \(M_{\rm tot}=M_{\rm p}+M_{\rm s}\) is the total black-hole mass and \(r_{\rm ps}\) is the orbital separation. We examine three physically distinct cases, each applying different corrections to the observed $\sim$5.26 yr modulations in $\gamma$-ray emissions:

\begin{enumerate}
\item \textit{Redshift correction:}  
      Rest-frame period \(P_{\rm ps}=P_{\rm obs}/(1+z)=2.08\,\rm yr\).

\item \textit{Doppler boosting + redshift correction:}  
      Rest-frame precession period (using \(\Gamma=17.51^{+7.32}_{-6.76}\) from the precession+nutation model, which is consistent with \citep{robinson2024neutrino}):
      \begin{equation}
      P_{\rm prec}=\frac{\Gamma^2 P_{\rm obs}}{1+z}\approx600\,\rm yr \nonumber
      \end{equation}
      .

\item \textit{Orbital-driven helical (non-ballistic) motion:}  
      \begin{equation}
      P_{\rm prec}=\frac{P_{\rm obs}}{(1+z)(1-\beta\cos\Omega\cos\phi)}\approx28\,\rm yr, \nonumber
      \end{equation}
      where \(\Omega\) and \(\phi\) are taken from Tables~\ref{tab:jet_precession} and~\ref{tab:jet_precession_nutation}.
\end{enumerate}

The binary orbital plane is inclined by an angle $\theta$ relative to the accretion disk, so the jet precesses inside a cone of half-opening angle $\Omega=\theta$. We adopt $\Omega\approx 30^{\circ}$; independent MCMC fits yield $\Omega={37.81^{+7.90}_{-8.62}}^{\circ}$ (pure precession) and $\Omega={29.36^{+7.62}_{-4.82}}^{\circ}$ (precession+nutation).

The outer radius (in units of PC) of the rigidly precessing disk is given by
\begin{align}
    r_{\rm{d}}^{\rm{out}} = \left[ \frac{8\pi}{3} \left( \frac{5-n}{7-2n}\right) \frac{(1+z)}{P^{'} \rm{cos}\Omega} \frac{r_{\rm{ps}}^3}{\sqrt{G M_{\rm{tot}}}} \right]^{2/3} \frac{x_{\rm{p}}^{1/3}}{(1-x_{\rm{p}})^{2/3}} ,
    \label{eq:rout_SMBBH}
\end{align}

where $x_{\rm p}=M_{\rm p}/M_{\rm tot}$, $n=3/2$ (non-relativistic gas) or $n=3$ (relativistic gas), and the condition $r_{\rm d}^{\rm out}<r_{\rm ps}$ must hold. We fix $M_{\rm p}=4\times10^8\,M_\odot$ \citep{robinson2024neutrino} and explore separations $0.001<r_{\rm ps}<0.1\,\rm pc$.

If such systems emit gravitational waves, the loss of orbital energy causes the binary separation to decrease and the orbital period to shorten progressively, ultimately leading to a merger. The corresponding merger timescale can be expressed as \citep{shapiro1983physics}:

\begin{align}
t_{\rm merger} &= 3.65\times10^5 
\left(\frac{r_{\rm init}}{0.01\,\rm pc}\right)^4
\left(\frac{M_{\rm tot}}{4\times10^8\,M_\odot}\right)^{-3} \nonumber \\
&\quad \times 
\left(\frac{x_{\rm p}}{0.5}\right)^{-1}
\left(\frac{x_{\rm s}}{0.5}\right)^{-1}
\,\rm yr
\label{eq:tmerger}
\end{align}

\begin{figure*}
    \centering
    \includegraphics[width=0.7\textwidth]{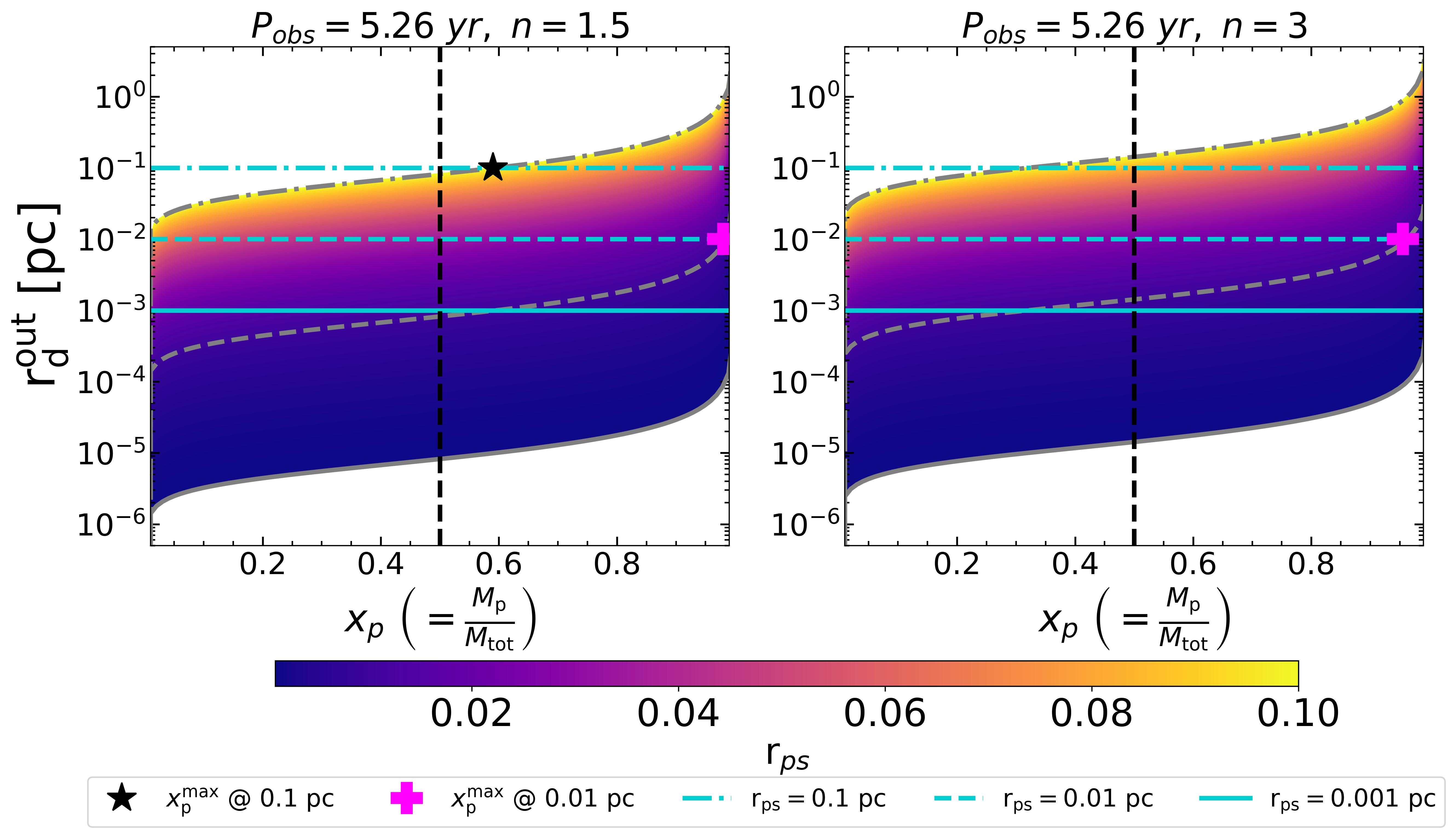}
    \caption{Precessing outer disk radius as a function of the primary-to-total mass ratio, $X_{\rm p} = M_{\rm p}/M_{\rm tot}$, for different polytropic index ($n$) values of gas. In this analysis, the precession period is corrected for Doppler boosting and cosmological redshift effects. The vertical black dashed line indicates $x_{\rm p} = 0.5$. The horizontal yellow lines correspond to binary orbital separations of $r_{\rm ps} = 0.001$, 0.01, and 0.1 pc, representing the characteristic orbital radii of the binary system. The color bar denotes the binary separation, spanning the range $0.001~{\rm pc} < r_{\rm ps} < 0.1~{\rm pc}$. The precessing outer disk radius is computed assuming an orbital inclination of $i = 30^{\circ}$, as inferred from the jet-precession model fit. The black $"*"$ and magenta $"+"$ symbols indicate the maximum primary mass ratio, $x_{\rm p}^{\rm max}$, corresponding to binary separations of $r_{\rm ps}$=0.1 pc and 0.01 pc, respectively. }
    \label{fig-LT_Precession_Binary_sys}
\end{figure*}

In case of Doppler$+$redshift corrected period, in a birnary system, for a relatively large separation of $r_{\rm ps}=0.1\,\mathrm{pc}$, we found a $x_{\rm p}^{\rm max}$ to be 0.59 for $n=3/2$, leading to a merger time scale of $t_{\rm merger}$ of order of $\sim 10^8$ yr. In contrast, for a small BH component separation of $r_{\rm ps}=0.01\,\mathrm{pc}$, $x_{\rm p}^{\rm max}$ approaches unity ($x_{\rm p}^{\rm max}$=0.98 for $n=3/2$), resulting in a merger time scale of $t_{\rm merger}\sim 10^5$ yr. Such high merger time scales are compatible with the long-term variability observed in PKS 1424-41. Similarly, in relativistic case ($n=3$), a separation of $r_{\rm ps}=0.01\,\mathrm{pc}$ yields a $x_{\rm p}^{\rm max}$ of 0.955, corresponding to a $t_{\rm merger}\sim 10^5$ yr, see Fig~\ref{fig-LT_Precession_Binary_sys} and Tab~\ref{tab:t_merger}. For the remaining cases corresponding to different timescales correction, the results are not favored.

\begin{table}[htbp]
\setlength{\extrarowheight}{6pt}
\setlength{\tabcolsep}{0.2pt}
\centering
\caption{Constraints on the parameters of a binary system in which a non-relativistic accretion disk undergoes precession with a period of $P$=5.26 yr.}
\label{tab:t_merger}
\begin{tabular}{lcccccc}
\hline
 & \(r_{\rm ps}\) (pc) & \(x_{\rm p}^{\rm max}\) & \(M_{\rm p}^{\rm max}\) (\(M_\odot\)) & \(x_{\rm s}^{\rm min}\) & \(M_{\rm s}^{\rm max}\) (\(M_\odot\)) & \(t_{\rm merger}\) (yr) \\
\hline
\multicolumn{7}{c}{{Redshift correction}} \\
\(n=3/2\) & 0.01 & 0.83 & \(6.64\times10^8\) & 0.17 & \(1.36\times10^8\) & \(\sim8\) \\
\(n=3\)   & 0.01 & 0.655 & \(5.24\times10^8\) & 0.345 & \(2.76\times10^8\) & \(\sim5\) \\
\hline
\multicolumn{7}{c}{{Doppler boosting + redshift}} \\
\(n=3/2\) & 0.1  & 0.59 & \(4.72\times10^8\) & 0.41 & \(3.28\times10^8\) & \(\sim4\times10^8\) \\
\(n=3/2\) & 0.01 & 0.98 & \(7.84\times10^8\) & 0.02 & \(1.60\times10^7\) & \(\sim5.8\times10^5\) \\
\(n=3\)   & 0.01 & 0.955 & \(7.64\times10^8\) & 0.045 & \(3.6\times10^7\) & \(\sim2.6\times10^5\) \\
\hline
\multicolumn{7}{c}{{Non-ballistic}} \\
\(n=3/2\) & 0.001 & 0.965 & \(7.72\times10^8\) & 0.035 & \(2.8\times10^7\) & \(\sim270\) \\
\(n=3\)   & 0.001 & 0.93  & \(7.44\times10^8\) & 0.069 & \(5.6\times10^7\) & \(\sim140\) \\
\hline
\end{tabular}
\end{table}

\begin{figure}
    \centering
    \includegraphics[width=0.4\textwidth]{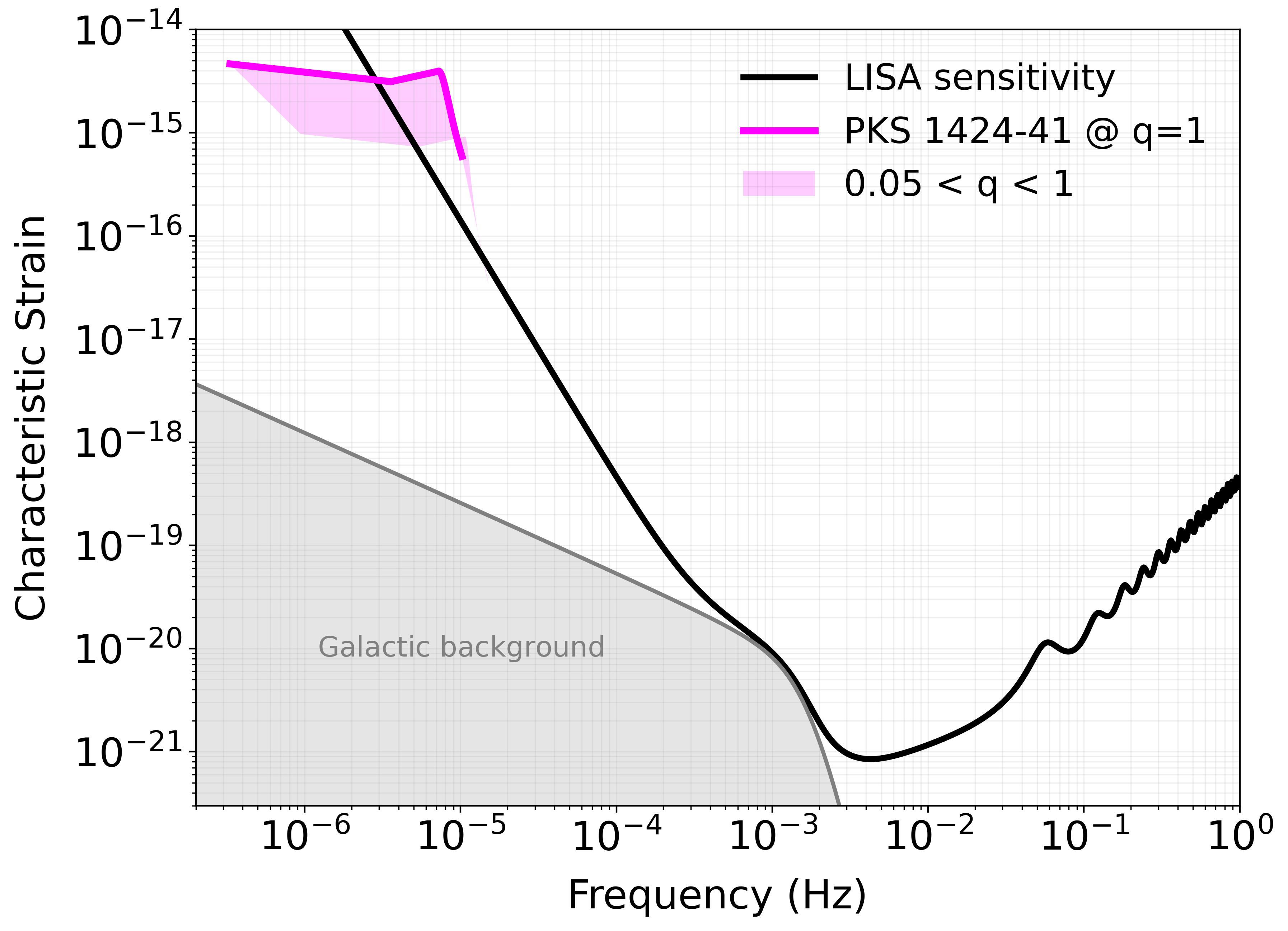}
    \caption{Prospects for the detection of a supermassive binary black hole by \textit{LISA} for a system with a primary black hole mass of $4 \times 10^{8}\,M_{\odot}$, evaluated at four years prior to coalescence. The black curve shows the expected \textit{LISA} sensitivity for a nominal 4-yr mission. The magenta curve represents the gravitational-wave signal from a binary supermassive black hole at $z = 1.522$ with an equal mass ratio ($q = 1$), starting four years before coalescence and evolving through the inspiral, merger, and ringdown phases. The shaded magenta region indicates the range of gravitational-wave signals corresponding to mass ratios $0.05 < q < 1$. The grey shaded region denotes the galactic confusion background.}
    \label{fig-LISA}
\end{figure}

\section{Prospect of Gravitational-Wave detection with LISA}\label{sec:lisa}

The characteristic gravitational-wave (GW) strain produced by a circular supermassive black hole binary (SMBHB) in the quadrupole approximation, as measured by NANOGrav, is given by \citep{aggarwal2019nanograv, arzoumanian2021nanograv}
\begin{equation}
h_{\rm GW}
=
\frac{2\left(G\,\mathcal{M}_{\oplus}\right)^{5/3}
\left(\pi f_{{\rm GW},\oplus}\right)^{2/3}}
{c^{4}\,D_{L}},
\label{eq:PTA_strain}
\end{equation}

where $\mathcal{M}_{\oplus} = \left(M_{1}M_{2}\right)^{3/5}/\left(M_{1}+M_{2}\right)^{1/5}$ is the chirp mass in the observer frame, expressed in terms of the component black hole masses $M_{1}$ and $M_{2}$ \citep{maggiore2008gravitational}. Here, $D_{L}$\footnote{Throughout this work, we adopt a luminosity distance $D_{\rm L}$ = 11.29 Gpc \citep{adachi2012analytical} by following these parameters$:$ at the source redshift of $z = 1.522$, with cosmological parameters corresponding to a $\Lambda$CDM of the Universe with $\Omega_{\rm m}$ = 0.29, $\Omega_{\lambda}$ = 0.7, and $H_0 = 69.6 \ \rm km \ \rm s^{-1} \ \rm Mpc^{-1}$.}=11.2 Gpc denotes the luminosity distance to the source, $f_{{\rm GW},\oplus}$ is the gravitational-wave frequency measured by the observer, and $G$ and $c$ are the gravitational constant and the speed of light, respectively.

Pulsar Timing Arrays (PTAs) are sensitive to nanohertz gravitational waves by monitoring the pulse arrival times of an ensemble of highly stable millisecond pulsars over decade-long baselines. Their primary targets are the stochastic gravitational-wave background (GWB), produced by the superposition of numerous unresolved supermassive black hole binaries (SMBHBs), and continuous gravitational waves (CWs) emitted by individual nearby SMBHBs in slowly evolving, nearly circular orbits. PTAs are most sensitive to binaries with total masses of $10^8$-$10^{10}$ M$_{\odot}$, orbital periods ranging from months to decades, and gravitational-wave frequencies of a few to several hundred nanohertz. The NANOGrav 11-year analysis found no significant evidence for individual CW sources and placed a sky-averaged 95$\%$ upper limit on the strain amplitude of h$_0 <7.3 \times 10^{-15}$ at the most sensitive frequency of $\sim$8 n$Hz$ \citep{aggarwal2019nanograv, arzoumanian2021nanograv}.

The chirp mass in the source rest frame is related to the observed chirp mass by $\mathcal{M} = \mathcal{M}_{\oplus}/(1+z)$, 
where $z$ is the redshift of the source. The observed gravitational-wave frequency is related to the observed orbital period $P_{\oplus}$ of the SMBBH system as $f_{{\rm GW}\oplus} = \frac{2}{P_{\oplus}}$. For an observed orbital period of $P_{\oplus} \simeq 5.26~{\rm yr}$, the corresponding gravitational-wave frequency in the observer frame is
$f_{{\rm GW}\oplus} \simeq 1.20\times10^{-8}\ {\rm Hz}$,
while the rest-frame gravitational-wave frequency is
$f_{\rm GW} = f_{{\rm GW}\oplus}/(1+z) \simeq 4.77\times10^{-9}\ {\rm Hz}$. We estimate the GW strain detectable by Pulsar Timing Arrays (PTAs) for a range of binary mass ratios, $0.05 < q < 1$, using Eq.~\ref{eq:PTA_strain}, \citep{o2022unanticipated}. Over this range, the predicted strain amplitude lies in the range $3.4\times10^{-20} \;\lesssim\; h_{\rm GW} \;\lesssim\; 5.5\times10^{-19}$, corresponding to observer-frame chirp masses $6.56\times10^{7}\,M_{\odot} \;\lesssim\; \mathcal{M}_{\oplus} \;\lesssim\; 3.48\times10^{8}\,M_{\odot}$. The resulting strain amplitudes are well below the sensitivity thresholds of current PTA experiments for individual source, implying that this system is not detectable as a single binary with present PTA observations \citep{zhu2014all}.
 
We further investigate the measurement capabilities of the future space-based gravitational-wave observatory Laser Interferometer Space Antenna (LISA), focusing on its sensitivity to the detection and characterization of massive black hole binaries. Given that LISA is expected to detect gravitational waves from binary systems with total masses in the range $\sim10^4 - 10^{9} \ \rm{M}_{\odot}$, see \citep{robson2019construction, katz1997precessing, liao2021discovery} for more details on LISA parameter estimation. We estimate the expected signal-to-noise ratio (SNR) for our source of interest from LISA sensitivity curve using phenomenological waveform models that incorporate the inspiral, merger, and ringdown phases of the coalescence. Fig.~\ref{fig-LISA} illustrates that the candidate binary supermassive black hole system is expected to eventually merge within the LISA frequency band. For a range of component masses and mass ratios, the resulting merger signals would be detectable by LISA over a nominal 4-yr mission, achieving a signal-to-noise (SNR) ratio of $\rm SNR\sim9$ at a redshift of $z=1.522$. We further estimate the gravitational-wave strain for mass ratios in the range $ 0.05<q<1$ and compare these predictions with the LISA sensitivity curve, see Fig~\ref{fig-LISA}.

\section{Summary and conclusion}\label{sec:summary}
\begin{enumerate}
\item First direct evidence of compound jet dynamics: $\gamma$-ray emissions reveals two superposed QPOs with periods $5.26 \pm 0.72\,\mathrm{yr}$, $0.96 \pm 0.07\,\mathrm{yr}$, and $\sim 1.3 \pm 0.23$ yr. The local and global significances associated with longer period are $4.26\sigma$ and $2.96\sigma$, respectively, while the shorter timescale QPOs exhibit a significance exceeding 3$\sigma$. Jet-precession + nutation modeling (MCMC) simultaneously reproduces both periodicities with best-fit parameters $P_p \approx 1907$ days, $P_n \approx 350$ days, $\Gamma \approx 17.5$, $\Omega_p \approx 29.4^\circ$, $\Omega_n \approx 1.1^\circ$, confirming wobbling and precessing jet motion in PKS\,1424--41.

\item The persistent $\sim$5.26 yr QPO is most naturally explained by orbital-driven jet precession in a binary supermassive black hole (BSBH) system with primary mass $M_p \approx 4 \times 10^8\,M_\odot$. Doppler-boosting + redshift-corrected scenarios yield realistic merger timescales of $10^5$--$10^8$ yr for separations 0.01--0.1 pc. Single black-hole Lense--Thirring precession is disfavored. The shorter $\sim$0.96 yr QPO arises from intrinsic jet nutation/wobbling.

\item If PKS\,1424--41 hosts a sub-parsec SMBH binary, it is a prime multi-messenger target for LISA. The system is expected to merge within the LISA frequency band, producing a detectable gravitational-wave signal (SNR $\sim 9$ for a nominal 4-yr mission at mass ratios $0.05 < q < 1$). The predicted strain lies below current PTA sensitivity but well within LISA's reach, establishing this blazar as a key source for probing supermassive black hole binaries.
\end{enumerate}

\section{ACKNOWLEDGMENT}
\begin{acknowledgments}
A. Sharma is grateful to Prof. Sakuntala Chatterjee at S.N. Bose National Centre for Basic Sciences, for providing the necessary support to conduct this research.
\end{acknowledgments}

%

\vspace{5mm}
\facilities{Fermi-LAT}








\bibliography{sample631}{}
\bibliographystyle{aasjournal}



\end{document}